\documentclass[11pt]{article}
\usepackage{latexsym}
\usepackage{graphicx}
\usepackage{amsmath}
\usepackage{amssymb}
\usepackage{amsfonts}
\usepackage{epsfig}
\usepackage{fancybox}
\usepackage{natbib}
\usepackage{array}
\usepackage{subfig}
\usepackage{url}
\usepackage{natbib}
\usepackage{listings}
\usepackage{algpseudocode}
\algdef{SxnE}[FOR]{For}{EndFor}[1]{\algorithmicfor\ #1 }

\graphicspath{{figure/}}
\allowdisplaybreaks[3]

\newtheorem{remark} {\bf Remark}

\newcommand{\bcen}{\begin{center}}
\newcommand{\ecen}{\end{center}}

\newcommand{\beqn}{\begin{equation}}
\newcommand{\eeqn}{\end{equation}}
\newcommand{\beqns}{\begin{equation*}}
\newcommand{\eeqns}{\end{equation*}}

\newcommand{\beqnary}{\begin{eqnarray}}
\newcommand{\eeqnary}{\end{eqnarray}}
\newcommand{\beqnarys}{\begin{eqnarray*}}
\newcommand{\eeqnarys}{\end{eqnarray*}}

\newcommand{\bary}{\begin{array}}
\newcommand{\eary}{\end{array}}

\newcommand{\benm}{\begin{enumerate}}
\newcommand{\eenm}{\end{enumerate}}

\newcommand{\bitem}{\begin{itemize}}
\newcommand{\eitem}{\end{itemize}}

\newcommand{\mbR}{\mathbb{R}}

\def\cA{{\cal A}}

\def\cN{{\cal N}}

\newcommand{\bx}{{\bf x}}

\newcommand{\bzero}{\mathbf{0}}
\newcommand{\bbeta}{\mbox{\boldmath{$\beta$}}}
\newcommand{\tbbeta}{\tilde{\mbox{\boldmath{$\beta$}}}}

\newcommand{\sbbeta}       {\mbox{\scriptsize\boldmath{$\beta$}\unboldmath}}

\newcommand{\overn}{\frac{1}{n}}

\newcommand{\pr}{\bold{P}}

\newcommand{\veps}{\varepsilon}

\begin{document}

\begin{center}
{\Large \bf A unified algorithm for the non-convex penalized estimation: The \textsf{ncpen} package }\\[10pt]
\vspace{0.3cm} \textrm{Dongshin Kim$^1$, Sangin Lee$^2$\footnote{D. Kim and S. Lee contributed equally to this work.} and Sunghoon Kown$^3$\footnote{Corresponding author: shkwon0522@gmail.com}} \\[0pt]
\vspace{0.1cm} \textit{Pepperdine University$^{1}$, Chungnam National University$^2$ and Konkuk University$^{3}$}\\[4pt]
\end{center}
\vspace{1mm}

\begin{abstract}
Various \textsf{R} packages have been developed for the non-convex penalized estimation
but they can only be applied to the smoothly clipped absolute deviation (SCAD) or minimax concave penalty (MCP).
We develop an R package, entitled \textsf{ncpen}, for the non-convex penalized estimation
in order to make data analysts to experience other non-convex penalties.
The package \textsf{ncpen} implements a unified algorithm based on the convex concave procedure and modified local quadratic approximation algorithm,
which can be applied to a broader range of non-convex penalties,
including the SCAD and MCP as special examples.
Many user-friendly functionalities such as generalized information criteria, cross-validation and $\ell_2$-stabilization are provided also.
\end{abstract}


\section{Introduction}\label{introduction}
The penalized estimation has been one of the most important statistical techniques for high dimensional data analysis,
and many penalties have been developed such as the least absolute shrinkage and selection operator (LASSO) \citep{tibshirani1996regression},
smoothly clipped absolute deviation (SCAD) \citep{fan2001variable}, and minimax concave penalty (MCP) \citep{zhang2010nearly}.

In the context of \textsf{R}, many authors released fast and stable \textsf{R} packages
for obtaining the whole solution path of the penalized estimator for the generalized linear model (GLM).
For example,
\textbf{lars} \citep{efron2004least},
\textbf{glmpath} \citep{park2007l1} and
\textbf{glmnet} \citep{friedman2007pathwise} implement the LASSO.
Packages such as
\textbf{plus} \citep{zhang2010nearly},
\textbf{sparsenet} \citep{mazumder2011sparsenet},
\textbf{cvplogit} \citep{jiang2014majorization} and
\textbf{ncvreg} \citep{breheny2011coordinate} implement the SCAD and MCP.
Among them, \textbf{glmnet} and \textbf{ncvreg} are very fast, stable, and well-organized,
presenting various user-friendly functionalities such as the cross-validation and $\ell_2$-stabilization \citep{zou2005regularization,huang2013balancing}.

The non-convex penalized estimation has been studied by many researchers
\citep{fan2001variable,kim2008smoothly,huang2008asymptotic,zou2008one,zhang2012general,kwon2012large,friedman2012fast}.
However, there is still a lack in research on the algorithms that exactly implement the non-convex penalized estimators for the non-convexity of the objective function.
One nice approach is using the coordinate descent (CD) algorithm \citep{tseng2001convergence,breheny2011coordinate}.
The CD algorithm fits quite well for some quadratic non-convex penalties
such as the SCAD and MC \citep{mazumder2011sparsenet,breheny2011coordinate,jiang2014majorization}
since each coordinate update in the CD algorithm becomes an easy convex optimization problem with a closed form solution.
This is the main reason for the preference of the CD algorithm implemented in many \textsf{R} packages such as \textbf{sparsenet} and \textbf{ncvreg}.
However, coordinate updates in the CD algorithm require extra univariate optimizations
for other non-convex penalties such as the log and bridge penalties \citep{zou2008one,huang2008asymptotic,friedman2012fast},
which severely lowers the convergence speed.
Another subtle point is that the CD algorithm requires standardization of the input variables
and need to enlarge the concave scale parameter in the penalty \citep{breheny2011coordinate}
to obtain the local convergence, which may cause to lose an advantage of non-convex penalized estimation \citep{kim2012global}
and give much different variable selection performance \citep{lee2015stand}.

In this paper, we develop an \textsf{R} package \textsf{ncpen} for the non-convex penalized estimation
based on the convex-concave procedure (CCCP) or difference-convex (DC) algorithm \citep{kim2008smoothly,shen2012likelihood}
and the modified local quadratic approximation algorithm (MLQA) \citep{lee2016modified}.
The main contribution of the package \textsf{ncpen} is that it encompasses most of existing non-convex penalties, including
the truncated $\ell_1$ \citep{shen2013constrained},
log \citep{zou2008one,friedman2012fast},
bridge \citep{huang2008asymptotic},
moderately clipped LASSO \citep{kwon2015moderately},
sparse ridge \citep{huang2013balancing,choi2013sparse}
penalties as well as the SCAD and MCP
and covers a broader range of regression models: multinomial and Cox models as well as the GLM.
Further, \textsf{ncpen} provides two unique options:
the investigation of initial dependent solution paths and non-standardization of input variables, which allow the users more flexibility.

The rest of the paper is organized as follows.
Section 2 describes the algorithm implemented in \textsf{ncpen} with major steps and details.
Section 3 and 4 introduces various options in \textsf{ncepn} with numerical illustrations.
The paper concludes with remarks.

\section{An algorithm for the non-convex penalized estimation}
We consider the problem of minimizing
\beqn\label{ncp:pro}
    Q_{\lambda}(\bbeta) = L(\bbeta) + \sum_{j=1}^pJ_{\lambda}(|\beta_j|),
\eeqn
where $\bbeta=(\beta_1,\ldots,\beta_p)^T$ is a $p$-dimensional parameter vector of interest, $L$ is a convex loss function
and $J_\lambda$ is a non-convex penalty with tuning parameter $\lambda>0$.
We first introduce the CCCP-MLQA algorithm for minimizing $Q_\lambda$ when $\lambda$ is fixed,
and then explain how to construct the whole solution path over a decreasing sequence of $\lambda$s by using the algorithm.

\subsection{A class of non-convex penalties}
We consider a class of non-convex penalties that satisfy $J_\lambda(|t|) = \int_0^{|t|} \nabla J_\lambda (s)ds, t\in\mbR$
for some non-decreasing function $\nabla J_\lambda$ and
\beqn\label{pen:decomp}
    D_\lambda(t) = J_\lambda(|t|)-\kappa_\lambda |t|
\eeqn
 is concave function, where $\kappa_\lambda  = \lim_{t\to0+}\nabla J_\lambda(t)$.
The class includes most of existing non-convex penalties:
SCAD \citep{fan2001variable},
\beqns
    \nabla J_{\lambda}(t) = \lambda I[0<t<\lambda] + \{(\tau\lambda-t)/(\tau-1)\}I[\lambda \leq t<\tau\lambda]
\eeqns
for $\tau>2$, MCP \citep{zhang2010nearly},
\beqns\label{pen:mcp}
     \nabla J_{\lambda}(t) = (\lambda-t/\tau)I[0< t<\tau\lambda]
\eeqns
for $\tau>1$, truncated $\ell_1$-penalty \citep{shen2013constrained},
\beqns \label{pen:tlp}
    \nabla J_{\lambda}(t) = \lambda I[0<t<\tau]
\eeqns
for $\tau>0$, moderately clipped LASSO \citep{kwon2015moderately},
\beqns \label{pen:tlp}
    \nabla J_{\lambda}(t) = (\lambda-t/\tau)[0<t<\tau(\lambda-\gamma)] + \gamma [t\geq \tau(\lambda-\gamma)]
\eeqns
for $\tau>1$ and $0\leq\gamma\leq \lambda$,
sparse ridge \citep{choi2013sparse},
\beqns \label{pen:tlp}
    \nabla J_{\lambda}(t) = (\lambda-t/\tau)I[0<t<\tau\lambda/(\tau\gamma+1)] + \gamma t I[t\geq\tau\lambda/(\tau\gamma+1)]
\eeqns
for $\tau>1$ and $\gamma\geq0$,
modified log \citep{Zou2005}.
\beqns \label{pen:tlp}
    \nabla J_{\lambda}(t) = (\lambda/\tau)[0<t<\tau] + (\lambda/t)[t\geq\tau]
\eeqns
for $\tau>0$, and modified bridge \citep{huang2008asymptotic}
\beqnarys \label{pen:tlp}
    \nabla J_{\lambda}(t) = (\lambda/2\sqrt{\tau})[0<t<\tau] + (\lambda/2\sqrt{t})[t\geq\tau]
\eeqnarys
for $\tau>0$.

The moderately clipped LASSO and sparse ridge are simple smooth interpolations between the MCP (near the origin) and the LASSO and ridge, respectively.
The log and bridge penalties are modified to be linear over $t\in(0,\tau]$ so that they have finite right derivative at the origin.
See the plot for graphical comparison of the penalties introduced here.

\begin{center}
    \begin{figure}
        \begin{center} \includegraphics[width=9cm,angle=270]{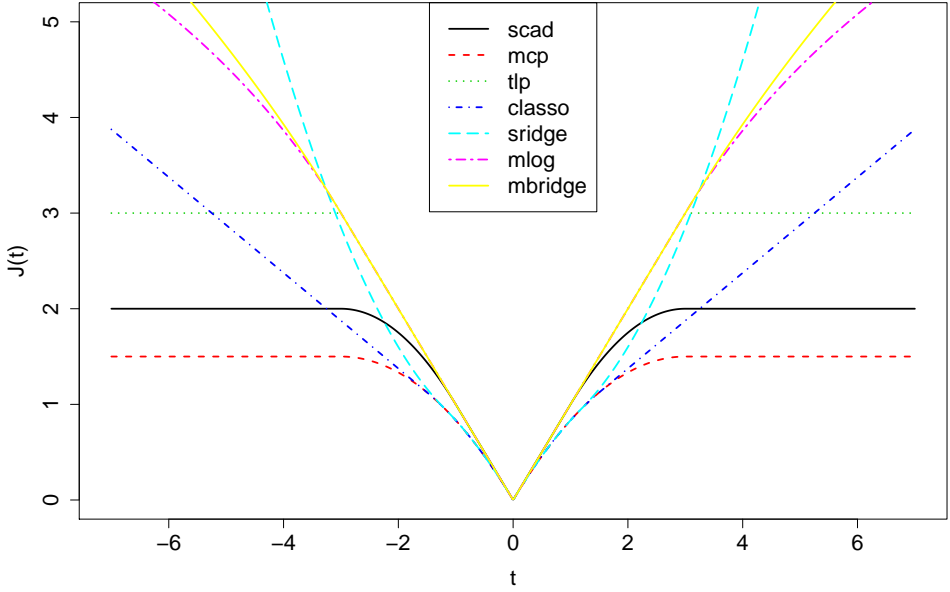}
        \end{center}
            \caption{Plot of various penalties with $\lambda=1, \tau=3$ and $\gamma=0.5$.}
        \label{fig:sub}
    \end{figure}
\end{center}

\subsection{CCCP-MLQA algorithm}
The CCCP-MLQA algorithm iteratively conducts two main steps:
CCCP \citep{yuille:rangarajan:cccp:2003} and MLQA \citep{lee2016modified} steps.
The CCCP step decomposes the penalty $J_\lambda$ as in \eqref{pen:decomp}
and then minimizes the tight convex upper bound obtained from a linear approximation of $D_\lambda$.
The MLQA step first minimizes a quadratic approximation of the loss $L$
and then modifies the solution to keep descent property.

\subsubsection{Concave-convex procedure}
The objective function $Q_\lambda$ in \eqref{ncp:pro} can be rewritten by using the decomposition in \eqref{pen:decomp} as
\beqn\label{obj:decomp}
    Q_{\lambda}(\bbeta) = L(\bbeta) + \sum_{j=1}^p  D_{\lambda}(\beta_j) + \kappa_\lambda \sum_{j=1}^p  |\beta_j|
\eeqn
so that $Q_\lambda(\bbeta)$ becomes a sum of convex, $L(\bbeta)+ \kappa_\lambda \sum_{j=1}^p  |\beta_j|$,
and concave, $\sum_{j=1}^p  D_{\lambda}(\beta_j)$, functions.
Hence the tight convex upper bound of $Q_\lambda(\bbeta)$ \citep{yuille:rangarajan:cccp:2003} becomes
\beqn\label{obj:cccp}
    U_{\lambda}(\bbeta;\tbbeta) = L(\bbeta) + \sum_{j=1}^p  \partial D_{\lambda}(\tilde\beta_j)\beta_j
    + \kappa_\lambda \sum_{j=1}^p |\beta_j|,
\eeqn
where $\tilde \bbeta=(\tilde\beta_1,\dots,\tilde\beta_p)^T$ is a given point and
$\partial D_\lambda(\tilde\beta_j)$ is a subgradient of $D_\lambda(\beta_j)$ at $\beta_j=\tilde\beta_j$.
Algorithm 1 summarizes the CCCP step for minimizing $Q_\lambda$.

\medskip

\noindent{\bf Algorithm 1: minimizing $Q_\lambda(\bbeta)$}\\
$~~~$ 1. Set $\tbbeta$.\\
$~~~$ 2. Update $\tbbeta$ by $\tbbeta = \arg\min_{\sbbeta}U_\lambda(\bbeta;\tbbeta)$.\\
$~~~$ 3. Repeat the Step 2 until convergence.



\subsubsection{Modified Local quadratic approximation}
Algorithm 1 includes minimizing $U_{\lambda}(\bbeta;\tbbeta)$ in \eqref{obj:cccp} by using a given solution $\tilde\bbeta$.
An easy way is iteratively minimizing local quadratic approximation (LQA) of $L$ around $\tilde\bbeta$:
\beqns
    \tilde L(\bbeta;\tbbeta) = L(\tbbeta)+\nabla L(\tbbeta)^T(\bbeta-\tbbeta)+(\bbeta-\tbbeta)^T\nabla^2L(\tbbeta)(\bbeta-\tbbeta)/2,
\eeqns
where $\nabla L(\bbeta)=\partial L(\bbeta)/\partial \bbeta$ and $\nabla^2 L(\bbeta)=\partial^2 L(\bbeta)/\partial \bbeta^2$.
Then $U_\lambda(\bbeta;\tbbeta)$ can be minimized by iteratively minimizing
\beqn\label{obj:lqa}
    \tilde U_{\lambda}(\bbeta;\tbbeta) = \tilde L(\bbeta;\tbbeta) + \sum_{j=1}^p  \partial D_{\lambda}(|\tilde\beta_j|)\beta_j
    + \kappa_\lambda \sum_{j=1}^p  |\beta_j|.
\eeqn
It is easy to minimize $\tilde U_{\lambda}(\bbeta;\tbbeta)$ since it is simply a quadratic function and the penalty term is the LASSO.
For the algorithm, we use the coordinate descent algorithm introduced by \cite{friedman2007pathwise}.
Note that the LQA algorithm may not have a descent property.
Hence, we incorporate the modification step to ensure the convergence of the LQA algorithm \citep{lee2016modified}.

\medskip

\noindent{\bf Algorithm 2: minimizing $U_{\lambda}(\bbeta;\tbbeta)$}\\
$~~~$ 1. Set $\tbbeta$.\\
$~~~$ 2. Find $\tbbeta^a = \arg\min_{\sbbeta}\tilde U_\lambda(\bbeta;\tbbeta)$.\\
$~~~$ 3. Find $\hat h = \arg\min_{h>0}  U_\lambda(h\tbbeta^a+(1-h)\tbbeta;\tbbeta)$.\\
$~~~$ 4. Update $\tbbeta$ by $\hat h\tbbeta^a+(1-\hat h)\tbbeta$.\\
$~~~$ 5. Repeat the Step 2--4 until convergence.

\subsection{Efficient path construction over $\lambda$}
Usually, the computation time of the algorithm rapidly increases
as the number of non-zero parameters increases or $\lambda$ decreases toward zero.
To accelerate the algorithm, we incorporate the active-set-control procedure
while constructing the solution path over a decreasing sequence of $\lambda$.

Assume that $\lambda$ is given and we have an initial solution $\tilde\bbeta$
which is expected to be very close to the minimizer of $Q_\lambda(\bbeta)$.
First we check the first order KKT optimality conditions:
\beqn\label{KKT}
    \partial Q_{\lambda}(\tilde\bbeta)/\partial\beta_j=0,j\in\cA ~~\mbox{and}~~
    |\partial Q_{\lambda}(\tilde\bbeta)/\partial\beta_j|\leq \kappa_{\lambda},j\in\cN,
\eeqn
where $\cA=\{j:\tilde\beta_j\neq0\}$ and $\cN=\{j:\tilde\beta_j=0\}$.
We stop the algorithm if the conditions are satisfied
else update $\cN$ and $\tilde\bbeta$ by $\cN=\cN \setminus\{j_{\max}\}$ and
\beqn\label{small:Q}
    \tilde\bbeta = {\arg\min}_{\beta_j=0,j\in\cN} Q_{\lambda}(\bbeta),
\eeqn
respectively, where $j_{\max}={\arg\max}_{j\in\cN}|\partial Q_{\lambda}(\tilde\bbeta)/\partial\beta_j|$.
We keep these iterations until the KKT conditions in \eqref{KKT} are satisfied with $\tilde\bbeta$.
The key step is \eqref{small:Q}
which is easy and fast to obtain by using Algorithm 1 and 2
since the objective function only includes the parameters in $\cA\cup\{j_{\max}\}$.

\medskip

\noindent{\bf Algorithm 3: minimizing $Q_{\lambda}(\bbeta)$}\\
$~~~$ 1. Set $\tbbeta$.\\
$~~~$ 2. Set $\cA=\{j:\tilde\beta_j\neq0\}$ and $\cN=\{j:\tilde\beta_j=0\}$.\\
$~~~$ 3. Check whether
        $\partial Q_{\lambda}(\tilde\bbeta)/\partial\beta_j=0,j\in\cA ~\mbox{and}~
        |\partial Q_{\lambda}(\tilde\bbeta)/\partial\beta_j|\leq \kappa_{\lambda},j\in\cN$.\\
$~~~$ 4. Update $\cN$ by $\cN \setminus\{j_{\max}\}$, where $j_{\max}={\arg\max}_{j\in\cN}|\partial Q_{\lambda}(\tilde\bbeta)/\partial\beta_j|$.\\
$~~~$ 5. Update $\tbbeta$ by  $\tilde\bbeta = {\arg\min}_{\beta_j=0,j\in\cN} Q_{\lambda}(\bbeta)$.\\
$~~~$ 6. Repeat the Step 2--5 until the KKT conditions satisfy.

{\remark
The number of the variables that violate the KKT conditions could be large for some high-dimensional cases.
In this case, it may be inefficient to add only one variable $j_{\max}$ into $\cA$.
It would be more efficient to add more variables into $\cA$.
However, when the number variables added is too large, it also is inefficient.
With many experiences, we found that the algorithm would be efficient with 10 variables.}

\medskip

In practice, we want to approximate the whole solution path or surface of the minimizer $\hat\bbeta^\lambda$ as a function of $\lambda$.
For the purpose, we first construct a decreasing sequence $\lambda_{\max}=\lambda_0>\lambda_1>\cdots>\lambda_{n-1}>\lambda_n=\lambda_{\min}$
and then obtain the corresponding sequence of minimizers $\hat\bbeta^{\lambda_0},\ldots,\hat\bbeta^{\lambda_n}$.
In general, we start from the largest value $\lambda=\lambda_{\max}=\max_j|\partial\nabla L(\bzero)/\partial\beta_j|$
since the $p$-dimensional zero vector is the exact minimizer of $Q_\lambda(\bbeta)$ when $\lambda\geq\lambda_{\max}$.
Then we continue down to $\lambda=\lambda_{\min}=\epsilon\lambda_{\max}$, where $\epsilon$ is a predetermined ratio such as $\epsilon=0.01$.
Once we obtain the minimizer $\hat\bbeta^{\lambda_{k-1}}$ then it is easy to find $\hat\bbeta^{\lambda_k}$ by using $\hat\bbeta^{\lambda_{k-1}}$ as an initial solution in Algorithm 3, which is expected to be close to $\hat\bbeta^{\lambda_k}$ for a finely divided $\lambda$ sequence.
This scheme is called the {\em warm start strategy}, which makes the algorithm more stable and efficient \citep{friedman2010regularization}.

\section{The \textsf{R} package \textbf{ncpen}}

In this section, we introduce various options and user-friendly functions implemented in the package \textbf{ncpen}
for the users. Next section will illustrate how these options make a difference in data analysis through numerical examples.

\subsection{$\ell_2$-regularization}
The option \verb"alpha" in the main functions forces the algorithm to solve the following penalized problem with the $\ell_2$-regularization or ridge effect \citep{Zou2005}.
\beqn\label{obs:ridge}
    Q_{\lambda}(\bbeta) = \overn\sum_{i=1}^{n}\ell_i(\bbeta) + \alpha\sum_{j=1}^{p}J_{\lambda}(|\beta_j|) + (1-\alpha)\lambda\sum_{j=1}^{p}\beta_j^2,
\eeqn
where $\alpha\in[0,1]$ is the value from the option \verb"alpha", which is the mixing parameter between the penalties $J_{\lambda}$ and ridge.
The objective function in \eqref{obs:ridge} includes the elastic net \citep{Zou2005} when $J_{\lambda}(t)=\lambda t$ and Mnet \citep{huang2016mnet} when $J_{\lambda}(\cdot)$ is the MC penalty.
By controlling the option \verb"alpha", we can treat the problem with highly correlated variables, and it also makes the algorithm more stable.

\subsection{Observation and penalty weights}
We can give different weights for each observation and penalty by the options \verb"obs.weight" and \verb"pen.weight",
which provides the minimizer of
\beqn\label{obs:weight}
    Q_{\lambda}(\bbeta) = \sum_{i=1}^{n} d_i \ell_i(\bbeta) + \sum_{j=1}^p w_jJ_{\lambda}(|\beta_j|),
\eeqn
where $d_i$ is the weight for the $i$th observation and $w_j$ is the penalty weight for the $j$th variable.
For example, controlling observation weights is required for the linear regression model with heteroscedastic error variance.
Further, we can compute adaptive versions of penalized estimators by giving different penalty weights as in the adaptive LASSO \citep{zou2006adaptive}.

\subsection{Standardization}
It is common practice to standardize variables prior to fitting the penalized models, but one may opt not to.
Hence, we provide the option \verb"x.standardize" for flexible analysis.
The option \verb"x.standardize=TRUE" means that the algorithm solves the original penalized problem in \eqref{ncp:pro}, with the standardized (scaled) variables,
and then the resulting solution $\hat\beta_j$ is converted to the original scale by $\hat\beta_j / s_j$, where $s_j=\sum_{i=1}^{n}x_{ij}^2/n$.
When the penalty $J_{\lambda}$ is the LASSO penalty, this procedure is equivalent to solving following penalized problem
\beqns
    Q^{s}_{\lambda}(\bbeta) = L(\bbeta) + \sum_{j=1}^p\lambda_j|\beta_j|,
\eeqns
where $\lambda_j=\lambda s_j$, which is another adaptive version of the LASSO being different from the adaptive LASSO \citep{zou2006adaptive}.

\subsection{Initial value}
We introduced the warm start strategy for speed up the algorithm, but the solution path, in fact, depends on the initial solution of the CCCP algorithm because of the non-convexity.
The option \verb"local=TRUE" in \textbf{ncpen} provides the same initial value specified by the option \verb"local.initial" into each CCCP iterations for whole $\lambda$ values.
The use of the option \verb"local=TRUE" makes the algorithm slower
but the performance of the resulting estimator would be often improved
as provided a good initial such as the maximum likelihood estimator or LASSO.

\subsection{Tuning parameter selection in \textbf{ncpen}}
The package \textsf{ncpen} includes several user-friendly functions
such as \verb"cv.ncpen" that conducts the cross-validation to select an optimal tuning parameter $\lambda$.
In addition, the function \verb"gic.ncpen" calculates theoretically optimal $\lambda$ based on the generalized information criterion \citep{wang2007tuning,wang2009shrinkage,fan2013tuning,wang2013calibrating}.

\section{Numerical illustrations}

\subsection{Elapsed times}
We consider the linear and logistic regression models to calculate the total elapsed time for constructing the solution path over 100 $\lambda$ values:
\beqn\label{exm:kwon}
     y=\bx^T\bbeta^*+\veps  ~~\mbox{and}~~ \pr(y=1|\bx)=e^{\bx^T\sbbeta}/(1+e^{\bx^T\sbbeta})
\eeqn
where $\bx \sim N_p(\bzero,\Sigma)$ with $\Sigma_{jk}=0.5^{|j-k|}$, $\beta_j=1/j$ for $j,k=1,\cdots,p$ and $\veps \sim N(0,1)$.
The averaged elapsed times of \textsf{ncpen} in 100 random repetitions are summarized in Table \ref{tab:time:n} and \ref{tab:time:p} for various $n$ and $p$,
where the penalties are the SCAD, MCP, truncated $\ell_1$ (TLP), moderately clipped LASSO (CLASSO), sparse ridge (SR), modified bridge (MBR) and log (MLOG).
For comparison, we try \textsf{ncvreg} for the SCAD also.
The results show that all methods in \textsf{ncpen} are feasible for high-dimensional data.


\begin{table}[h]
\begin{center}
\caption{Elapsed times for constructing the entire solution path where $p=500$ and various $n$} \footnotesize
\begin{tabular}{lrrrrrrrrr}
\hline
Model      &\multicolumn{1}{c}{$n$}  &\multicolumn{1}{c}{\textsf{ncvreg}} &\multicolumn{1}{c}{SCAD}
           &\multicolumn{1}{c}{MCP}  &\multicolumn{1}{c}{TLP}             &\multicolumn{1}{c}{CLASSO}
           &\multicolumn{1}{c}{SR}   &\multicolumn{1}{c}{MBR} &\multicolumn{1}{c}{MLOG} \\
\hline
Linear     &200  &0.0226 &0.1277 &0.1971 &0.0333 &0.0696 &0.0618  &0.0620  &0.0476   \\
regression &400  &0.0329 &0.1082 &0.2031 &0.0662 &0.1041 &0.1025  &0.1160  &0.0919   \\
           &800  &0.0347 &0.1008 &0.1867 &0.0865 &0.0993 &0.1067  &0.1425  &0.1197   \\
           &1600 &0.0665 &0.2035 &0.3170 &0.1717 &0.1847 &0.1983  &0.2669  &0.2301   \\
           &3200 &0.1394 &0.4341 &0.6173 &0.3541 &0.3962 &0.4161  &0.5505  &0.4678   \\
           &6400 &0.2991 &0.9853 &1.2045 &0.7955 &0.8788 &0.9066  &1.2281  &1.0148   \\
\hline
Logistic   &200  &0.0565 &0.0454 &0.0400 &0.0391 &0.0148 &0.0160  &0.0379 &0.0411    \\
regression &400  &0.0787 &0.1113 &0.0971 &0.0747 &0.0556 &0.0608  &0.0969 &0.0808    \\
           &800  &0.0907 &0.1570 &0.1623 &0.1198 &0.0777 &0.1015  &0.1511 &0.1298    \\
           &1600 &0.1682 &0.2965 &0.3007 &0.2294 &0.1640 &0.2088  &0.3002 &0.2451    \\
           &3200 &0.3494 &0.6480 &0.6258 &0.4655 &0.3513 &0.4423  &0.6395 &0.5305    \\
           &6400 &0.7310 &1.4144 &1.3711 &1.0268 &0.8389 &1.0273  &1.4445 &1.1827    \\
\hline
\end{tabular}\label{tab:time:n}
\end{center}
\end{table}

\begin{table}[h!]
\begin{center}
\caption{Elapsed times for constructing the entire solution path where $n=500$ and various $p$} \footnotesize
\begin{tabular}{lrrrrrrrrr}
\hline
Model      &\multicolumn{1}{c}{$p$}  &\multicolumn{1}{c}{\textsf{ncvreg}} &\multicolumn{1}{c}{SCAD}
           &\multicolumn{1}{c}{MCP}  &\multicolumn{1}{c}{TLP}             &\multicolumn{1}{c}{CLASSO}
           &\multicolumn{1}{c}{SR}   &\multicolumn{1}{c}{MBR} &\multicolumn{1}{c}{MLOG} \\
\hline
Linear     &200  &0.0150 &0.0733 &0.2201 &0.0433 &0.0629 &0.0981  &0.0909 &0.0721   \\
regression &400  &0.0210 &0.0664 &0.1588 &0.0532 &0.0617 &0.0678  &0.0941 &0.0813   \\
           &800  &0.0538 &0.1650 &0.2172 &0.1107 &0.1505 &0.1457  &0.1750 &0.1383   \\
           &1600 &0.0945 &0.2703 &0.2946 &0.1793 &0.2253 &0.2221  &0.2672 &0.2045   \\
           &3200 &0.1769 &0.5071 &0.5032 &0.3379 &0.3972 &0.3986  &0.4801 &0.3684   \\
           &6400 &0.3439 &1.0781 &1.0228 &0.7366 &0.8001 &0.8207  &1.0210 &0.7830   \\
\hline
Logistic   &200  &0.0590 &0.1065 &0.1029 &0.0750 &0.0465 &0.0696  &0.0978 &0.0804   \\
regression &400  &0.0568 &0.1054 &0.1044 &0.0753 &0.0453 &0.0593  &0.0941 &0.0809   \\
           &800  &0.1076 &0.1555 &0.1349 &0.1103 &0.0873 &0.0934  &0.1423 &0.1163   \\
           &1600 &0.1327 &0.1944 &0.1591 &0.1419 &0.1122 &0.1151  &0.1842 &0.1460   \\
           &3200 &0.2073 &0.3120 &0.2529 &0.2382 &0.1885 &0.1948  &0.3055 &0.2415   \\
           &6400 &0.3843 &0.5893 &0.4792 &0.4646 &0.3539 &0.3576  &0.5978 &0.4677   \\
\hline
\end{tabular}\label{tab:time:p}
\end{center}
\end{table}

\subsection{Standardization effect}
\begin{center}
    \begin{figure}
        \begin{center} \includegraphics[width=\linewidth]{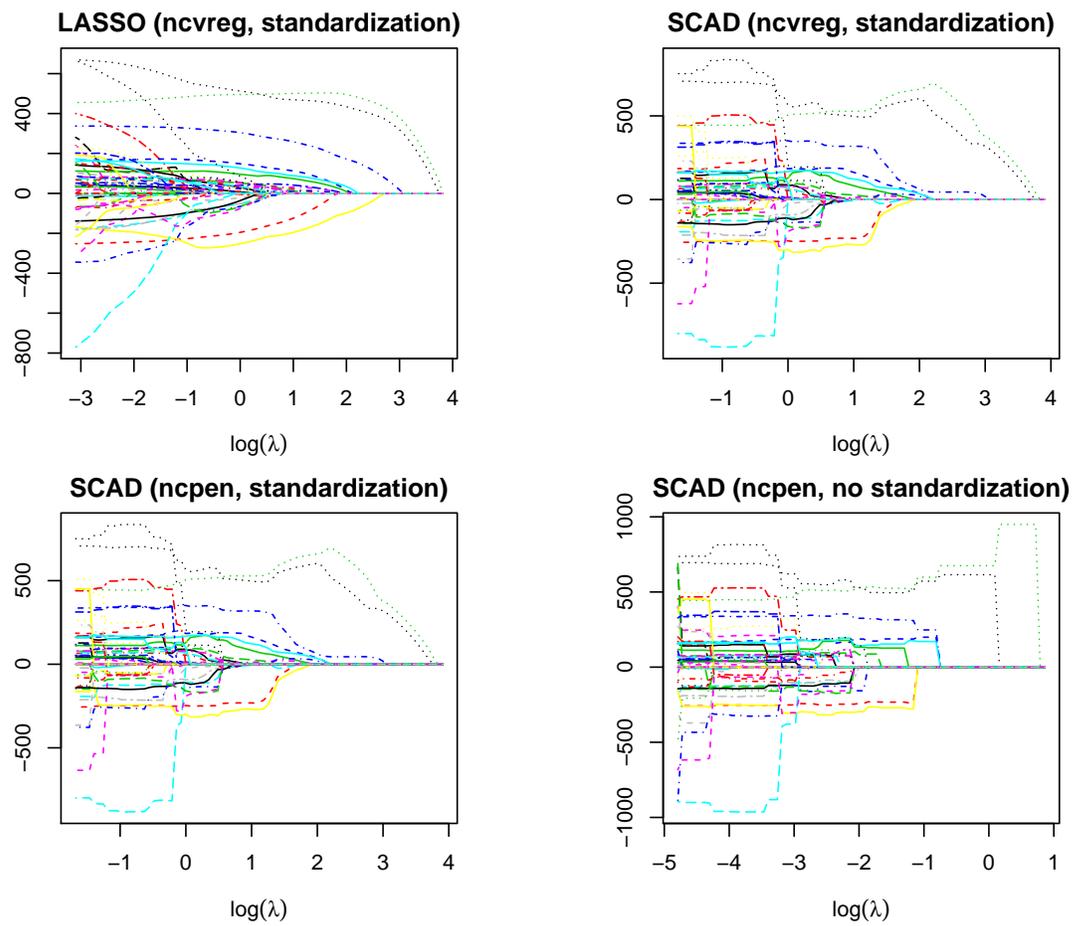}
        \end{center}
            \caption{Solution paths from the \textsf{ncvreg} and \textsf{ncpen} for the LASSO and SCAD}
        \label{fig:stand}
    \end{figure}
\end{center}

\begin{center}
    \begin{figure}
        \begin{center} \includegraphics[width=\linewidth]{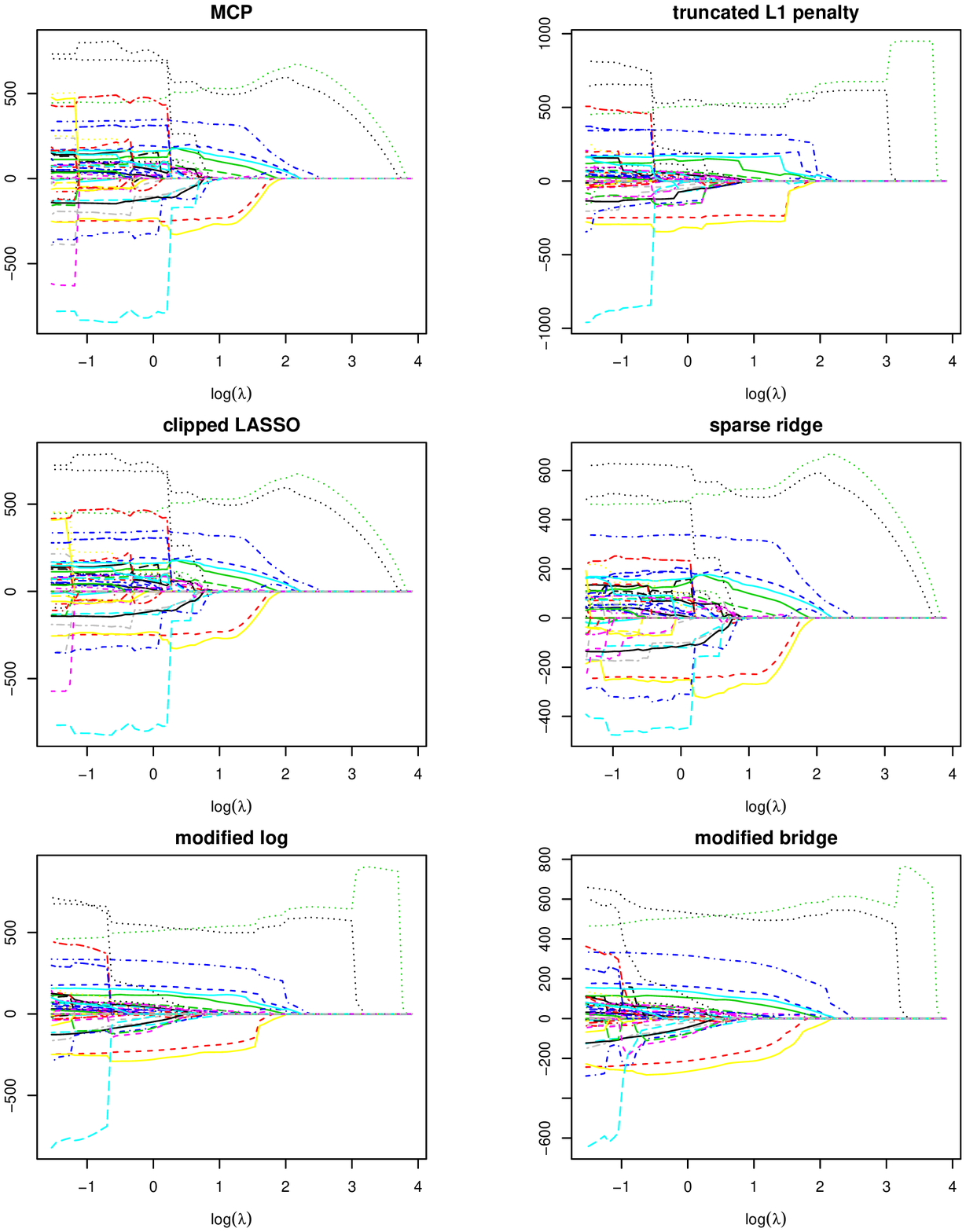}
        \end{center}
            \caption{Solution paths from \textsf{ncpen} with six non-convex penalties}
        \label{fig:other}
    \end{figure}
\end{center}

We compare the solution paths based on the diabetes samples available from \textsf{lars} package \citep{efron2004least},
where the sample size $n=442$ and the number of covariates $p=64$, including quadratic and interaction terms.
Figure \ref{fig:stand} shows four plots where the top two panels draw the solution paths from the LASSO and SCAD with $\tau=3.7$ given by \textsf{ncvreg}
and bottom two panels draw those from the SCAD with $\tau=3.7$ based on \textsf{ncpen} with and without standardization of covariates.
Two solution paths from \textsf{ncvreg} and \textsf{ncpen} with standardization are almost the same
since \textsf{ncvreg} standardizes the covariates by default, which is somewhat different from that of \textsf{ncpen} without standardization.
Figure \ref{fig:other} shows the solution paths from six penalties with standardization by default in \textsf{ncpen}:
the MCP, truncated $\ell_1$, modified log, bridge, moderately clipped LASSO and sparse ridge.

\subsection{$\ell_2$-regularization effect}
\begin{center}
    \begin{figure}
        \begin{center} \includegraphics[width=\linewidth]{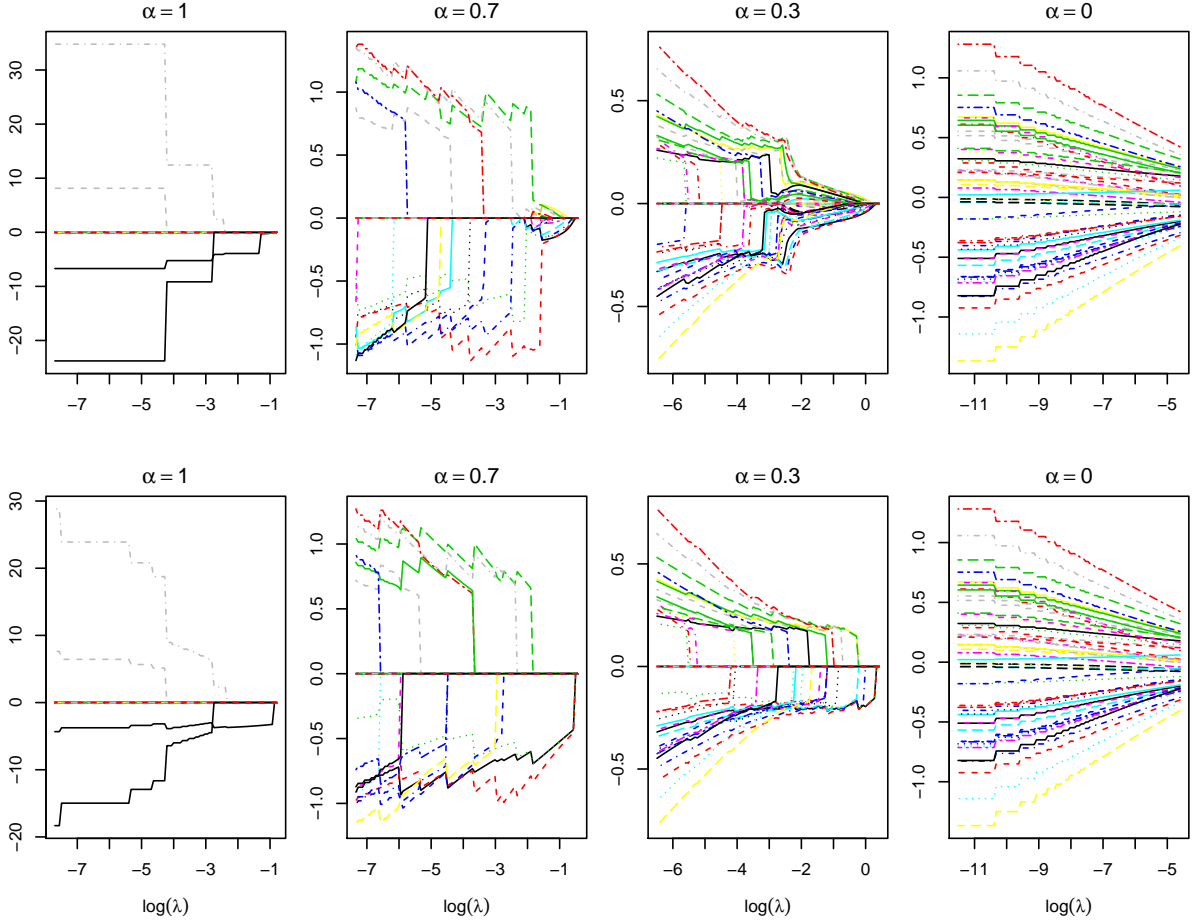}
        \end{center}
            \caption{Solution path traces with ridge penalty. Top and bottom panels are drawn from the SCAD and modified bridge penalties, respectively.}
        \label{fig:ridge}
    \end{figure}
\end{center}

There are cases when we need to introduce the ridge penalty for some reasons,
and \textsf{ncpen} provides a hybrid version of the penalties: $\alpha J_{\lambda}(|t|)+(1-\alpha)|t|^2$,
where $\alpha$ is the mixing parameter between the penalty $J_{\lambda}$ and ridge effects.
For example, the non-convex penalties often produce parameters
that diverge to infinity for the logistic regression because of perfect fitting.
Figure \ref{fig:ridge} shows the effects of ridge penalty
where the leukemia samples in \textsf{plsgenomics} are used for illustration.
The solution paths using the top 50 variables with high variances are drawn when $\alpha\in\{1,0.7,0.3,0\}$ for the SCAD and modified bridge penalties.
The solution paths without ridge effect ($\alpha=1$) tend to diverge as $\lambda$ decreases and become stabilized as the ridge effect increases ($\alpha\downarrow$ ).

\subsection{Initial based solution path}
\begin{center}
    \begin{figure}
        \begin{center} \includegraphics[width=\linewidth]{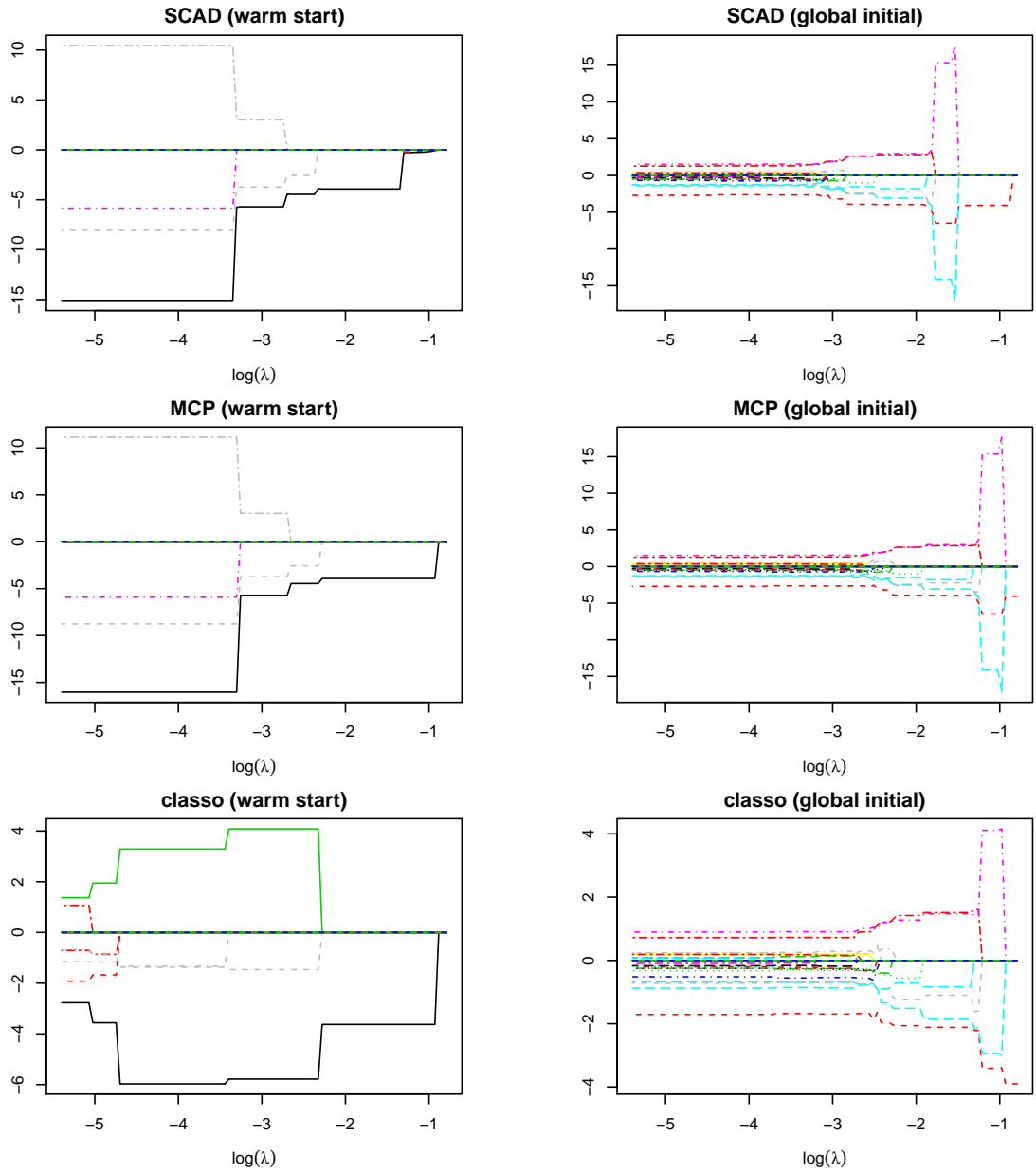}
        \end{center}
            \caption{Solution paths with warm start and global initial solution. Top and bottom panels are drawn from the SCAD and modified bridge penalties, respectively.}
        \label{fig:ini}
    \end{figure}
\end{center}

We introduced the warm start strategy for speed up the algorithm
but the solution path, in fact, depends on the initial solution because of the non-convexity.
For comparison, we use the leukemia samples and the results are displayed in Figure \ref{fig:ini} and Table \ref{tab:loc}.
In the figure, left panels show the solution paths for the SCAD, MCP and clipped LASSO obtained by the warm start,
and the right panels show those obtained by using the LASSO as a global initial for the CCCP algorithm.
Figure \ref{fig:ini} shows two strategies for initial provide very different solution paths, which may result in different performances of the estimators.
We compare the prediction accuracy and selectivity of the estimators by two strategies.
The results are obtained by 300 random partitions of data set divided into two parts, training (90\%) and test (10\%) datasets.
For each training data, the optimal tuning parameter values are selected by 10-fold cross-validation,
and then we compute the prediction error on each test datasets and the number of selected nonzero variables.
Table \ref{tab:loc} shows all methods by the global initial perform better than those by the warm start strategy.
In summary, the nonconvex penalized estimation depends on the initial solution,
and the non-convex penalized estimator by a good initial would improve its performance.


\begin{table}[h!]
\begin{center}
\caption{Comparison of the warm start and global initial strategies for each method.
The TNL and MIS represent the average of negative log-likelihood value and misclassification error on each test dataset,
and the NUM means the average of numbers of the selected variables on each training dataset.} \footnotesize
\begin{tabular}{crrrrrrrrrr}
\hline
&  \multicolumn{3}{c}{warm start} &&\multicolumn{3}{c}{global initial} \\
\cline{2-4}\cline{6-8}
Method &\multicolumn{1}{c}{TNL} &\multicolumn{1}{c}{MIS} &\multicolumn{1}{c}{NUM} &&\multicolumn{1}{c}{TNL} &\multicolumn{1}{c}{MIS} &\multicolumn{1}{c}{NUM} \\
\hline
SCAD   &0.4777(.0359) &0.0894(.0060) &1.61(.0502) &&0.1965(.0253) &0.0422(.0048) & 9.96(.3433)  \\
MCP    &0.4258(.0386) &0.1466(.0081) &0.81(.0307) &&0.2244(.0640) &0.0422(.0050) &10.28(.3337)  \\
TLP    &0.5242(.0956) &0.1666(.0082) &0.87(.0410) &&0.1264(.0186) &0.0327(.0041) &19.08(.1156)  \\
CLASSO &0.3370(.0268) &0.0905(.0060) &5.09(.1406) &&0.1330(.0152) &0.0466(.0046) & 6.39(.1143)  \\
SR     &0.1800(.0176) &0.0588(.0050) &8.76(.0857) &&0.0952(.0119) &0.0233(.0036) &16.99(.1687)  \\
MBR    &0.3997(.0305) &0.1105(.0070) &1.13(.0290) &&0.1804(.0262) &0.0427(.0045) & 6.84(.1004)  \\
MLOG   &0.5176(.0939) &0.1555(.0078) &0.95(.0408) &&0.1474(.0215) &0.0350(.0042) &13.46(.1416)  \\
\hline
\end{tabular}\label{tab:loc}
\end{center}
\end{table}

\section{Concluding remarks}
We have developed the \textsf{R} package \textsf{ncpen} for estimating generalized linear models with various concave penalties.
The unified algorithm implemented in \textsf{ncpen} is flexible and efficient.
The package also provides various user-friendly functions and user-specific options for different penalized estimators.
The package is currently available with a general public license (GPL) from the Comprehensive R Archive Network at
\url{https://CRAN.R-project.org/package=ncpen}.
Our \textsf{ncpen} package implements internal optimization algorithms in \textsf{C++} benefiting from \textsf{Rcpp} package \citep{eddelbuettel2011rcpp}.

\section*{Acknowledgements}
This research is funded by Julian Virtue Professorship from Center for Applied Research at Pepperdine Graziadio Business School,
the National Research Foundation of Korea (NRF) funded by Korea government (No. 2017R1C1B2010113 and 2017R1D1A1B03031239).

\setlength{\bibsep}{0.0pt}
\bibliographystyle{ims}

\end{document}